\documentstyle[aps,twocolumn]{revtex}
\begin{document}

\title{
Weak-Localization in Chaotic versus Non-chaotic Cavities: \\
A Striking Difference in the Line Shape}

\author{
A. M. Chang, H. U. Baranger, L. N. Pfeiffer, and K. W. West}

\address{
AT\&T Bell Laboratories, 600 Mountain Avenue, Murray Hill NJ 07974-0636
\medskip\\
{\rm [Submitted to Phys. Rev. Lett., May 26, 1994] } \medskip \\
\parbox{14cm}{\rm
We report experimental evidence that chaotic and non-chaotic scattering
through ballistic cavities display distinct signatures in quantum transport.
In the case of non-chaotic cavities, we observe a linear decrease in the
average resistance with magnetic field
which contrasts markedly with a Lorentzian behavior for a chaotic cavity.
This difference in line-shape of the weak-localization peak is related to
the differing distribution of areas enclosed by electron trajectories.
In addition, periodic oscillations are observed
which are probably associated with the Aharonov-Bohm effect through a periodic
orbit within the cavities.
}}

\maketitle

\narrowtext

A central question for ``quantum chaos'' in open systems is whether
scattering from a system which classically exhibits chaotic
dynamics will give rise to observable quantum signatures distinct from a
non-chaotic system \cite{QMSca}.
Considerable attention has been focused on the connection between the
quantum $S$-matrix and the classical dynamics
\cite{QMSca,Smilan,Jal90,LewWei91,Oak92,Lai92,Bar93,Lin93}.
Recent theoretical studies indicate that
the quantum transmission probability, which is intimately related to
the conductance of a microstructure, directly reflects the classical behavior
\cite{Jal90,Oak92,Bar93,Lin93}.
In particular, in the chaotic case the statistical distribution
of the quantum transmission probability as a function of either the incident
momentum $k$ \cite{Smilan} or magnetic field $B$ \cite{Jal90}
reflects the existence of a single classical trapping time scale:
the power spectrum of the conductance fluctuations $G(k)$ or $G(B)$
decays via a single exponential for several decades.
In addition, the average transport should show a negative magneto-resistance
peak, known as weak-localization (WL), centered about $B=0$
{\it with a line shape in the form of a Lorentzian} \cite{Bar93}.
In direct contrast, in the non-chaotic (integrable) case, the lack of a
single time scale gives rise to a power law tail in the power spectrum
for large frequencies \cite{Jal90,Lai92,Bar93,Lin93} and
{\it a highly unusual linear line shape} for the weak-localization peak
\cite{Bar93}.

In this Letter, we focus on the WL line shape. We present experimental
evidence that the shape is strikingly different for transport through
a ballistic chaotic cavity in the form of a stadium versus a non-chaotic
cavity in the form of a circle.  Specifically, we observe a Lorentzian line
shape for the stadium and a {\it linear} line shape for the circle down to
$\sim 0.09 B_{1/2}$, where $B_{1/2}$ refers to the half width
at half maximum.  Our results are obtained in microstructures fabricated from
a very high quality ${\rm GaAs/Al_{x}Ga_{1-x}As}$ heterostructure crystal.
{\it For each type of cavity, 48 devices are measured at once to average
out the universal conductance fluctuations.}
We directly compare these results with numerical calculations of nominally
identical cavities; the good agreement between theory and experiment strongly
supports our view that the difference in line shape is caused by the
difference in the character of the classical dynamics.

We extend our basic observation in two different ways.
(1) As the temperature is lowered from 4.2 K to 50 mK, the WL in the stadium
remains Lorentzian while that in the circle evolves from a
Lorentzian to the linear shape below 400 mK.  This
suggests that as the phase-coherence length, $L_{\phi}$,
grows, longer trajectories which enclose larger area begin to contribute
to coherent-back-scattering at lower magnetic fields.
(2) We present evidence that stable periodic
orbits can exist in a ballistic cavity by observing a single periodicity
Aharonov-Bohm effect at the lowest temperatures in certain nominally
rectangular structures.

Previous experimental studies of ``quantum chaos'' in microstructures
have focused on the universal conductance fluctuations (UCF) and WL peak of
{\it individual} cavities \cite{Marcus,Kel94,MJBer94} or on the
magnetoresistance of antidot arrays \cite{Weiss93}.
Marcus, et al. \cite{Marcus} investigated the power spectrum of $G(B)$ for a
stadium versus a circle: the fact that their circles showed more power at the
higher frequencies than their stadia was the first experimental evidence for
a difference between chaotic and non-chaotic behavior in quantum transport
(power law versus exponential decay).
The deviation occurred only after the power had decayed substantially,
however,
so that effects of elastic and inelastic scattering have to be carefully
considered. In fact,
the power spectrum for the circle can be fit by an exponential over one
and a half decades: the resulting rate is smaller than for the stadium but
is consistent with a reasonable estimate of the small-angle mean-free-path
\cite{Lin93}. The possibility of differing interpretations of this seminal
work makes a clear observation of
the difference between chaotic and non-chaotic cavities essential.
Later studies along similar lines concentrated on the magnetotransport of an
individual stadium cavity \cite{Kel94,MJBer94} and therefore did not address
the issue of chaotic versus non-chaotic behavior. Finally, several studies
\cite{Marcus,Kel94,Weiss93}, particularly that on the magnetoresistance of
antidot arrays \cite{Weiss93}, have seen evidence for interference effects
between short non-universal paths, such as the circular trajectories
enclosing an antidot in the case of Ref. \cite{Weiss93}.
We return to this issue at the end of the paper.

Our devices are fabricated on a doping well ${\rm GaAs/Al_{x}Ga_{1-x}As}$
heterostructure crystal with an electron density of
$3.3 \times 10^{11}$ cm$^{-2}$ and a mobility of
$1.8 \times 10^6$ cm$^{2}$/Vs.
Measurement is performed with a lock-in amplifier at 23 Hz at a current of
20 nA for $T \leq 1.6$ K and 100 nA at $T \geq 2.4$ K.
{\it Length scales:} The transport mean free path is 17 $\mu$m
compared to a typical cavity diameter of $\leq 1.25$ $\mu$m.  The small-angle
scattering length is estimated from the observed 2\%
modulation of $G(B)$ (at $T$= 0.5 K) due to electron focusing \cite{HvH88}
for injection and
collection point contacts separated by 11.6 $\mu$m in linear distance.
Assuming an exponential decay in the number of unperturbed electrons as a
function of distance,
we deduce a small-angle scattering length of 4 $\mu$m.  The phase-coherence
length, $L_{\phi}$, is estimated to exceed
15 $\mu$m below 400 mK \cite{KurChang92}.
{\it Size:} The cavities are
fabricated by electron beam lithography and ECR (low voltage reactive ion)
etching.  Both the stadium and circle cavities are fabricated in one
electron beam write on the same device chip. The perimeter is defined via
a $1500 {\rm \AA} $ wide line etched to depletion.  The
lithographic dimension of the stadium as shown in the inset of Fig. 1(a) is
$1.25 \times 0.85 $ $\mu$m with a 0.4 $\mu$m straight portion.
The circle of equivalent area is of diameter 1.08 $\mu$m [Fig. 1(b)].
The electron gas resides $568 {\rm \AA} $ below the surface, and the
etching is $200 {\rm \AA} $ in depth.  From the pinch-off
characteristic of narrow constrictions $\approx 1000 {\rm \AA} $
wide, we estimate a depletion length of $300 \pm 100 {\rm \AA} $,
yielding an area of 0.81 $\mu$m$^{2}$ with a corresponding magnetic field
scale of 51.4 Gauss for one flux quantum penetrating the enclosed area.
{\it Geometry:} The ratio of the entrance plus exit widths
to the perimeter is roughly 1 to 6.5.  Therefore, the number of phase
coherent bounces before exiting is cut off mostly by the openings rather than
$L_{\phi}$ when $T \leq 400$ mK.  The device for a given cavity type
consists of 48 cavities arranged in 6 rows in series of 8 in parallel.
The spacing between rows is 25 $\mu$m, comfortably larger than $L_{\phi}$.

In Fig. 1 we show the weak-localization, negative magneto-resistance peak
for the stadium and circle samples at 50 mK.  The
resistance value refers to that of a single cavity.  The stadium WL peak can
readily be fitted by a Lorentzian line shape with a peak height
$\Delta G$ of 0.4 $e^2 /h$ and a $B_{1/2}$ of 13 $\pm$ 1 Gauss.
In contrast, the line shape for the circle is certainly non-Lorentzian,
following an unusual linear decrease with magnetic field.  The peak is
characterized by $\Delta G$$= 0.22$ $e^2 /h$ and $B_{1/2} = 11 \pm 1$ Gauss.
While we concentrate on studying the lineshape, we note
that the magnitude of the WL peak in the stadium is consistent with recent
theoretical work \cite{Bar94Jal94} which predicts a universal magnitude of
0.25 $e^2/h$ per spin channel in the fully chaotic case.

The possibility of such an unusual line shape in a non-chaotic
(integrable) system was first pointed out in the work of Baranger, et
al. \cite{Bar93} who gave a general semiclassical argument connecting
the linear line shape to the existence of a power-law distribution of
classical areas in a non-chaotic system.  In order to demonstrate the
connection between our experimental results and the theory,
in Fig. 2 we show the change in conductance [$- \Delta G$=$G(B$=0)$-G(B)$]
obtained from numerical calculations for ballistic billiards which have
the same nominal shape as the experimental structures.
The conductance is calculated
through its relation to the total transmission, $G = (e^2/h)T$,
by using the recursive Green function method to obtain $T$ for a
discretized billiard (useing $ka \approx 1.4$) \cite{Bar91}. The average
conductance needed to make contact with the experiments is found by averaging
over energy in the range for which there are  2-9 propagating modes in the
leads.

Note the clearly Lorentzian line shape for the stadium and the more
triangular shape for the circle. The resemblance between experiment and
theory is remarkable for both cavities. Since only the difference in shape
between the chaotic stadium and the non-chaotic circle influences the
calculation, the different WL line shapes are clearly connected to the
difference in classical dynamics.
{\it We believe these results in combination provide strong evidence that
chaotic and non-chaotic scattering in ballistic cavities indeed give rise
to the experimentally observed difference in transport.}

An important issue in interpreting the experimental results is the
influence of small-angle scattering. The first quantitative theoretical
treatment was given by Lin, Delos, and Jensen \cite{Lin93}:
they emphasize that small-angle scattering in a circle changes the classical
distribution of areas from a power-law to an exponential, but one whose
characteristic area is different from that of a stadium with the same
geometric area.
Thus one must seriously consider the effects of any disorder even
when the transport mean-free-path is large.
We have carried out quantum calculations for the stadium
and circle billiards in the presence of a smooth disordered potential. The
disorder is formed by choosing a random value for the potential
at every fifth lattice site (within a range $[-W_{dis} /2, W_{dis} /2]$) and
linearly interpolating in between \cite{Bar90}. The transport mean-free-path
for this potential is 10-20 times larger than the
total mean-free-path.
Fig. 2(b) shows that the linear WL line shape in the non-chaotic
case is not destroyed by a smooth disordered potential whose strength is
chosen so as to match the total and transport mean-free-paths in the
experimental structures ($W_{dis} = 0.25$).
For comparison we show that strong boundary roughness scattering
does change the theoretical line shape to a Lorentzian, as expected.
In this case, $W_{dis} = 5$ was used on the last meshpoint before the
hard wall yielding diffusive scattering from the surface.
We conclude that small-angle scattering is weak enough in these
experiments so that the non-chaotic nature of the classical paths in the
ideal circular cavity is actually observed.

Conclusive evidence that phase-coherence is essential for the experimental
observations is provided by the temperature evolution of the WL peak shown
in Fig. 3. For the stadium, the line shape is Lorentzian for the entire
temperature range, as shown for the Lorentzian fits
at 50 mK and 1.6 K.  On the other hand, while
the circle appears to be Lorentzian-like at the higher temperatures, albeit
with a slightly cusped peak at $B=0$, the full linear behavior develops below
$\approx 400$ mK.  It appears that the longer trajectories which enclose
larger areas and contribute to the negative magneto-resistance
at the smaller B fields become sufficiently phase coherent only at these
lower temperatures.

While the general WL lineshape discussed above is related to the full
distribution of areas of classical paths, semiclassical
theory also suggests that interference between short non-universal paths
should produce magnetotransport effects particular to certain shapes.
Indeed, some experimental evidence for the role of small area trajectories
has already been reported \cite{Marcus,Kel94,Weiss93}.
In Fig. 4, we present particularly clear evidence for the existence
of a stable periodic orbit in a nominally rectangular cavity of
$1.2 \times 0.75$ $\mu$m: the magneto-resistance shows pronounced
modulation as a function of B at low temperatures $\leq 400$ mK.
The Fourier power spectrum exhibits double peaks at 14 and 16 cycles/kG.
The average period of 67 Gauss corresponds to an area of
0.62 $\mu$m$^{2}$ for the penetration of one flux quantum compared to
to estimated area of 0.79 $\mu$m$^{2}$ for the cavity .  The inset shows
a possible periodic orbit.
Because this result is obtained in a device containing 48 cavities, we
believe this is strong evidence for the existence of a stable orbit
common to a significant fraction of all cavities.

{\it Acknowledgements---} We thank D. J. Bishop for support and continued
interest in the course of this work.

\begin{figure}
\caption{
The magneto-resistance for (a) 48 stadium cavities, and (b) 48
circle cavities at $T=$ 50 mK.  The weak localization peak line shape shows
a Lorentzian behavior for the chaotic, stadium cavities.  In contrast,
the line shape for the non-chaotic, circle cavities shows a highly unusual,
triangular shape (linearly decreasing).
The resistance value is normalized to a single cavity.  The
vertical bar indicates the equivalent change in conductance, $ \Delta G$.
Insets show electron micrographs of the cavities which are fabricated on
a high quality ${\rm GaAs/Al_{x}Ga_{1-x}As}$ heterostructure crystal.
}
\label{wow}
\end{figure}

\begin{figure}
\caption{
Calculated magnetoconductance ($\times -1$) as a function of flux through the
geometric area of the cavity for the (a) stadium and (b) circle
shown as insets. The lineshape is Lorentzian for the
ballistic stadium (solid squares) as well as for
a stadium with strong surface roughness scattering (diamonds).
The lineshape is more triangular for both the ballistic circle
(triangles) and the circle with a weak smooth disordered potential
(solid squares), but changes to Lorentzian for strong surface roughness
scattering (diamonds). For the disordered potential, the total mean-free-path
is approximately 5 times the diameter of the cavity.
The similarity of lineshape between this calculation and the experiment
(Fig. 1) is striking for both structures.
}
\label{theory}
\end{figure}

\begin{figure}
\caption{
The temperature evolution of the magneto-resistance for the
(a) stadium cavities, and (b) circle cavities.
{}From top to bottom, $T$= 50 mK, 200 mK, 400 mK, 800 mK,
1.6 K, 2.4 K, and 4.2 K.  The dash-dotted lines are Lorentzian fits.
For the stadium cavities, the weak localization line shape is
Lorentzian at all temperatures.  For the circle cavities, the line shape
is Lorentzian only at higher temperatures above 2.4 K.  The linearly
decreasing triangular line shape develops fully below 400 mK, showing
that phase-coherence is essential in producing this shape.
}
\label{Tdep}
\end{figure}

\begin{figure}
\caption{
(a) Magneto-resistance for 48 nominally rectangular cavities
normalized to a single cavity.  The
traces correspond to the temperatures of 50 mK (top), 200 mK, 400 mK,
800 mK, and 1.6 K (bottom), with each successive trace displaced
downward by 0.6 $ k \Omega $.  In addition to the weak localization peak at
$B$=0, periodic modulation of the magneto-resistance is present probably
caused by interference around a periodic orbit in the structure.  Panel (b)
shows that the Fourier power spectrum exhibits double peaks at 14 and 16
cycles/kG corresponding to an average period of 67 Gauss.  The inset to
panel (a) contains an electron micrograph of the cavity.  The inset to panel
(b) delineates a possible periodic orbit (dashed lines) inside the cavity
(solid curve) for which lateral depletion of the 2D electron gas from etching
has rounded the corners.
}
\label{per_orb}
\end{figure}

\end{document}